\documentclass{article}
\usepackage{amsfonts}

\usepackage{graphicx}
\usepackage{amsmath}

\newtheorem{theorem}{Theorem}
\newtheorem{acknowledgement}[theorem]{Acknowledgement}

\newtheorem{lemma}[theorem]{Lemma}

\newenvironment{proof}[1][Proof]{\textbf{#1.} }{\ \rule{0.5em}{0.5em}}
\input{tcilatex}

\begin{document}

\title{Low Rank Separable States Are A Set Of Measure Zero Within The Set of Low
Rank States}
\author{Robert Lockhart \\
Math Dept, USNA, Annapolis, MD, 21402\\
rbl@usna.edu}
\maketitle

\begin{abstract}
It is well known that the set of separable pure states is measure 0 in the
set of pure states. We extend this fact and show that the set of rank r
separable states is measure 0 in the set of rank r states provided r is less
than $\prod_{i=1}^{p}n_{i}+p-\sum n_{i}$.
\end{abstract}

Recently quite a few authors have looked at low rank separable and entangled
states. (See \cite{horo} and the references therein and \cite{hao} and \cite
{chen} Therefore it makes sense to determine the size of the set of rank $r$
separable states within the set of rank $r$ states. For rank 1, it is well
known that the separable states are a set of measure zero. This contrasts
with the maximal rank case, where the separable states not only are not
measure zero, but contain an open set.

The purpose of this note is to show that the rank 1 result is also true for
many other low ranks. In particular, suppose we have $p$ particles modelled
on the Hilbert space $\mathbb{C}^{n_{1}}\otimes \cdots \otimes \mathbb{C}%
^{n_{p}}$. Then the following is true.

\begin{theorem}
Let S$_{r}$ be the set of rank r separable matrices on $\mathbb{C}%
^{n_{1}}\otimes \cdots \otimes \mathbb{C}^{n_{p}}$ and D$_{r}$ the set of
all rank r density matrices. S$_{r}$ is measure 0 in D$_{r}$, for all $%
r<N+p-\sum n_{i}$, where $N=n_{1}\cdots n_{p}$
\end{theorem}

This is an extension of recent results of Chen. In \cite{chen}, he showed in
the bipartite case where $H=\mathbb{C}^{m}\otimes \mathbb{C}^{m}$ that $%
S_{r} $ is measure 0 in $D_{r}$ if $r<2m-3$. As a result of Theorem 1, we
see that $S_{r}$ is in fact measure 0 in $D_{r}$ if $r<m^{2}-2m+2=(m-1)^{2}.$
Since the number of possible ranks is $m^{2}$, we see that in this case the
proportion of those $r$ for which $S_{r}$ is measure 0 in $D_{r}$ is $%
(1-(1/m^{2}))^{2}.$ In the case of p-qubits, $S_{r}$ is measure 0 in $D_{r}$
if $r<2^{p}-2p+p=2^{p}-p.$ Since the number of possible ranks in this case
is $2^{p}$, we see that the proportion of those $r$ for which $S_{r}$ is
measure 0 in $D_{r}$ is $1-(p/2^{p})$. Thus we see that the ranks of ''low''
rank matrices can be quite high.

The proof will use Sard's Theorem \cite{milnor}to show the set of ranges of
separable rank $r$ density matrices is measure 0 in the set of ranges of
rank $r$ density matrices, i.e. within the set of $r$-dimensional subspaces
of $\mathbb{C}^{N}$; i.e. within $G(N,r)$, the Grassmann manifold of $r-$%
planes in $\mathbb{C}^{N}$ \cite{dieu}, if $r<N+p-\sum n_{i}$. That this is
what we need to consider follows from the fact $D_{r}$ is $G(N,r)\times
Herm_{1}^{+}(r)$, where $Herm_{1}^{+}(r)$ is the space of Hermitian $r\times
r$ matrices that are positive definite and trace 1. For those not familiar
with Sard's Theorem, we will need only a simple consequence of it. Namely,
we need the fact that if $f:M\rightarrow N$ is a smooth function (i.e.
infinitely differentiable function) between an $m$-dimensional space $M$ and
an $n-$dimensional space $N$ and if $m<n$, then $f(M)$ is measure zero in $N$%
. This should be intuitively obvious: the \ smooth image of a lower
dimensional space in a larger one is measure zero. Think for instance of a
smooth curve in a plane. The curve has 0 area.

\begin{lemma}
Suppose A and B are positive semi-definite linear operators. Then $%
Ker(A+B)=KerA\cap KerB$ and $Range(A+B)=RangeA+RangeB$.
\end{lemma}

\begin{proof}
Clearly $KerA\cap KerB\subset Ker(A+B)$ and $Range(A+B)\subset
RangeA+RangeB. $ Suppose $v\in KerA\cap KerB$. Then $0=\left\langle
(A+B)v,v\right\rangle =\left\langle Av,v\right\rangle +\left\langle
Bv,v\right\rangle $. Since $A$ and $B$ are positive semi-definite, this
means $Av=0$ and $Bv=0,$ hence $KerA\cap KerB=Ker(A+B).$ Since $Ker(A+B)$ is
the orthogonal complement of $Range(A+B),$ it follows that $KerA\cap KerB$
is the orthogonal complement of $Range(A+B).$ But $KerA\cap KerB$ is the
orthogonal complement of $Range(A+B) $, so $Range(A+B)=RangeA+RangeB$.
\end{proof}

If $A$ is a separable density matrix, then $A$ is the convex combination of
projections onto product states. It follows from the lemma that the range of 
$A$ therefore has a basis of product states \cite{horod}. The product states
are precisely the image of $\mathbb{P}^{n_{1}-1}\times \cdots \times $ $%
\mathbb{P}^{n_{p}-1}$in $\mathbb{P}^{N-1}$ under the map induced by tensor
product on $\mathbb{C}^{n_{1}}\times \cdots \times \mathbb{C}^{n_{p}}$,
where $\mathbb{P}^{k}$ is $k$-dimensional complex projective space (i.e. the
space of rays in $\mathbb{C}^{k+1}$).

As for the manifold of $r$-dimensional subspaces of $\mathbb{C}^{N},$ i.e.,
the Grassmann manifold, $G(N,r)$, it is obtained by first considering $%
\mathbb{C}^{Nr}$ as being the set of $N\times r$ complex matrices and taking 
$L(N,r)$ to be the open subset consisting of those matrices with rank $r$. $%
G(N,r)$ is then the orbit space $GL(r,$ $\mathbb{C})\backslash L(N,r).$

Let $\widetilde{\mu }:\mathbb{P}^{n_{1}-1}\times \cdots \times $ $\mathbb{P}%
^{n_{p}-1}\rightarrow $ $\mathbb{P}^{N-1}$be the map induced by tensor
product and take $\mu :(\mathbb{P}^{n_{1}-1}\times \cdots \times \mathbb{P}%
^{n_{p}-1})^{r}\rightarrow (\mathbb{P}^{N-1})^{r}$ to be $(\widetilde{\mu }%
,...,\widetilde{\mu })$. Let $Q=\mu ^{-1}(\widetilde{L(}N,r)$, where $%
\widetilde{L}(N,r)$ is the image of $L(N,r)$ in $\mathbb{P}^{r(N-1)}$. Let $%
\widetilde{\pi :}$ $\widetilde{L}(N,r)\rightarrow G(N,r)$ be the projection
induced by $\pi :L(N,r)\rightarrow G(N,r)$. Then $\widetilde{\pi }\circ \mu
:Q\rightarrow G(N,r)$ has as its range the $r-$dimensional subspaces that
have a basis of product states.

Since $\widetilde{\pi }\circ \mu :Q\rightarrow G(N,r)$ is a smooth map, the
theorem follows from Sard's theorem and the facts that $\dim G(N,r)=r(N-r)$
and $\dim (\mathbb{P}^{n_{1}-1}\times \cdots \times \mathbb{P}%
^{n_{p}-1})^{r}=r(\sum (n_{i}-1))=r(p-\sum n_{i}).$ Thus the $\dim Q<\dim
G(N,r)$ if $r(p-\sum n_{i})<r(N-r)$ that is if $r<N+P-\sum n_{i}$.

\begin{acknowledgement}
Part of this work was done at the Naval Research Laboratory, where the
author is a part time member of Michael Steiner's quantum information group
\end{acknowledgement}

\begin{acknowledgement}
This paper is a revision of an earlier one. The author thanks Professor Hao
Chen for pointing out to him a mistake in the proof in the original paper.
The mistake was such that the assertion of the main theorem had to be
changed from $S_{r}$ has measure 0 for all $r<N$ to what it is now.
\end{acknowledgement}

\end{document}